\documentclass[%
 reprint,
 amsmath,amssymb,
 aps,
]{revtex4-2}

\usepackage{graphicx}% Include figure files
\usepackage{epsfig} % Include eps files
\usepackage{tikz}
\usepackage{pgfplots}
\usepackage{dcolumn}% Align table columns on decimal point
\usepackage{bm}% bold math
\usepackage[colorlinks=true, allcolors=blue]{hyperref}
\usepackage{braket}
\usepackage{float}
\usepackage{chngcntr}

\usepackage{todonotes}
\usepackage[normalem]{ulem}

\definecolor{DEIred}{RGB}{155,0,20}
\definecolor{Green532}{RGB}{101, 255, 0}
\definecolor{Pantone}{RGB}{65,150,180}
\definecolor{DarkGreen}{RGB}{10, 110, 10}

\pgfmathdeclarefunction{gauss}{3}{%
  \pgfmathparse{#3/(#2*sqrt(2*pi))*exp(-((x-#1)^2)/(2*#2^2))}%
}

\begin{document}

\preprint{APS/123-QED}

\title{Certification of genuine time-bin and energy-time entanglement \\with integrated photonics}% Force line breaks with \\

\author{Francesco B. L. Santagiustina$^{1,2}$}
\author{Costantino Agnesi$^{2}$}
\author{Alvaro Alarcón$^{3}$}
\author{Adán~Cabello$^{4,5}$}
\author{Guilherme B. Xavier$^{3}$}
\author{Paolo Villoresi$^{1,2}$}
\author{Giuseppe Vallone$^{2,1,6}$}
 \email{giuseppe.vallone@pd.infn.it}

\affiliation{%
 $^{1}$Istituto Nazionale di Fisica Nucleare (INFN)—Sezione di Padova, Via Marzolo 8, 35131, Padova, Italy \\
 $^{2}$Dipartimento di Ingegneria dell'Informazione, Universit\`a degli Studi di Padova, via Gradenigo 6B, 35131 Padova, Italy \\
 $^{3}$Institutionen för Systemteknik, Linköpings Universitet, Linköping 581 83, Sweden\\
 $^{4}$Departamento de F\'{\i}sica Aplicada II, Universidad de Sevilla, E-41012 Sevilla, Spain\\
 $^{5}$Instituto Carlos~I de F\'{\i}sica Te\'orica y Computacional, Universidad de Sevilla, E-41012 Sevilla, 
Spain
\\
$^6$ Dipartimento di Fisica e Astronomia, Universit\`a degli Studi di Padova, via Marzolo 8, IT-35131 Padova, Italy
}%

\date{\today}% It is always \today, today,

\begin{abstract}
Time-bin (TB) and energy-time (ET) entanglement are crucial resources for long-distance quantum information processing. Recently, major efforts have been made to produce compact high-quality sources of TB/ET entangled photons based on solid-state integrated technologies. 
However, these attempts failed to close the so-called ``post-selection loophole''.
Here, we present an integrated photonic general Bell-test chip for genuine (i.e., free of the post-selection loophole) TB and ET entanglement certification. We report a violation of a Bell inequality by more than 10 standard deviations using our device based on the ``hug'' interferometric scheme. The experiment also demonstrates that the hug scheme, previously exploited for ET entanglement, can also be used for genuine TB entanglement.
\end{abstract}

\maketitle

%\tableofcontents

%%%%%%%%%%%%%%%%%%%%%%%%%%%%%%%%%%%%%%%%%%%%%%%%%%%%%%%%%%%%%%%%%%%%%%%%%%%%%%%%

%\section{\label{sec:intro}Introduction }
{\it Introduction.---}Entanglement is a crucial resource in quantum communication protocols \cite{Horodecki_entanglement}, including quantum key distribution \cite{Ekert91}, 
quantum teleportation \cite{Bennett_teleportation}, and quantum secret sharing \cite{Hillery_QSS}. Entanglement has also caused major debates \cite{EPR, Bohr}, mainly concerning whether quantum systems have a hidden set of predetermined instructions (the so-called hidden variables) before a measurement operation is performed. This was settled by Bell, who showed that the assumptions of realism and locality cannot be simultaneously satisfied when the results from a correlation test are above a certain threshold \cite{Bell_1964}.
The most widely used of these correlation tests is the Clauser-Horne-Shimony-Holt (CHSH) inequality \cite{CHSH}, designed for bipartite systems with dichotomic measurement outputs. Due to experimental imperfections, a 
number of local hidden-variable models (LHVMs) have been derived over the years, which forces one to make assumptions in order to guarantee the validity of the Bell test \cite{Larsson_2014}. Recently major advances were made in closing all major loopholes simultaneously in experiments \cite{Hensen_2015, Giustina_2015, Shalm_2015, Rosenfeld_2017}, generating the possibility of ultra-secure device-independent quantum communication systems \cite{Nadlinger_2022, Zhang_2022, Liu_2022}.
Energy-time (ET) entanglement is a robust form of photonic entanglement that arises from the energy-time relation when photon pairs are produced in a non-linear medium \cite{Flamini_2019}. First proposed by Franson in 1989 \cite{Franson_1989}, Bell tests on ET photon pairs have been widely used in many quantum communication schemes due to their robustness for long-distance propagation \cite{Flamini_2019}. Time-bin (TB) entanglement, a popular modification where the excitation pump laser is already prepared in a superposition of an early and a late time-bin creating photon pairs in well-defined times, was demonstrated in 1999 \cite{Brendel99}. A major issue with both TB and ET schemes is related to the discovery of local hidden-variable models (LHVM) which explain the violation due to the post-selection of detection events, thus requiring extra assumptions to trust the Bell test result \cite{Aerts_1999,jogenfors_2014}. This post-selection loophole was first removed using hyper-entangled states \cite{Strekalov_1996}, then exploiting a topologically different interferometric arrangement called the ``hug'' interferometer allowing genuine ET entanglement \cite{rossi_generation_2008,cabello_proposed_2009, lima_experimental_2010, cuevas_long-distance_2013, carvacho_postselection-loophole-free_2015}, and finally through the use of active optical switches for genuine TB entanglement production \cite{vedovato_postselection-loophole-free_2018}. Removing the post-selection loophole is highly relevant since it has been exploited to experimentally hack ET and TB entanglement-based quantum key distribution systems \cite{Jogenfors_2015}. Improved compatibility and stability requirements for TB and ET entanglement sources  \cite{Caspani_2017, Moody_2020, Wang_2021} have encouraged the use of novel techniques such as quantum dots \cite{Jayakumar_2014, Versteegh_2015, Prilmuller_2018, Gines_2021}, micro-ring resonators on integrated photonics \cite{Grassani_2015, Wakabayashi_2015, Reimer_2016, Mazeas_2016, Fujiwara_2017, Ma_2017, Samara_2019, Lu_2019, Oser_2020, Steiner_2021}, and integrated waveguides \cite{Takesue_2007, Harada_2008, Takesue_2014, Sarrafi_2014, Xiong_2015, Autebert_2016, Li_2017, Zhang:18, Zhao_2020}. However, the violation of Bell’s inequality in these works cannot be fully certified due to the post-selection loophole present from the use of Franson's scheme.

In this work, we design and demonstrate an integrated photonics chip for the certification of both TB and ET entanglement sources that is based on the hug interferometric scheme, which is not affected by the post-selection loophole. Our chip is based on a silicon nitride platform \cite{Triplex} and was designed to be inserted between a user's source and detectors. It contains the complete hug scheme including the required unbalanced Mach-Zehnder interferometers with thermal elements to apply the measurement operators for the Bell test. Our results show that a single photonic integrated chip (PIC) can be used to certify genuine ET and TB entanglement sources, providing a solid verification tool that any user could use, especially when dealing with uncharacterized or untrusted sources, a widely relevant topic in quantum communication networks.

%%%%%%%%%%%%%%%%%%%%%%%%%%%%%%%%%%%%%%%%%%%%%%%%%%%%%%%%%%%%%%%%%%%%%%%%%%%%%%%%

%\vskip0.5cm
{\it Genuine time-bin entanglement.---}This Letter presents, to our knowledge, the first use of the hug configuration with a TB entangled state instead of an ET source pumped in continuous wave (CW) mode. We will thus briefly describe here the main differences between the two cases, while a more detailed treatment is given in  \ref{sec:AppendixA} of the Appendix. Extending the use of the hug interferometer to the TB case allows a reduction in the frequency stability requirements of the pump laser and benefits from having specific photon generation times instead of a uniform distribution. This allows for synchronization of operations to be carried on the biphoton and eases the interference with other photons in more complex protocols.

We now briefly recall the difference between the Franson and hug configurations. The Franson scheme is based on the use of two independent (equally) unbalanced interferometers at the measurement stations. In the ET case, it is required that the imbalance is greater than the single photon coherence time but lower than the pump laser coherence time. Post-selection of coincident events should be applied: since the photons can be emitted at any time, there is no possibility of performing post-selection locally. In the TB case, the pump interferometer must have the same imbalance as the measurement interferometers. In this case, the three possible arrival times of each photon are known in advance. However, also in this case, to select only the central arrival times for both photons a non-local post-selection is necessary.

%%%%%%%%%%%%%%%%%%%%%%%%%%%%%%%%%%%%%%%%%%%%%%%%%%%%%%%%%%%%%%%%%%%%%%%%%%%%%%%%

\begin{figure}[h]
  \centering
  \includegraphics[width=\linewidth,trim=0cm 0cm 0cm 0cm,clip]{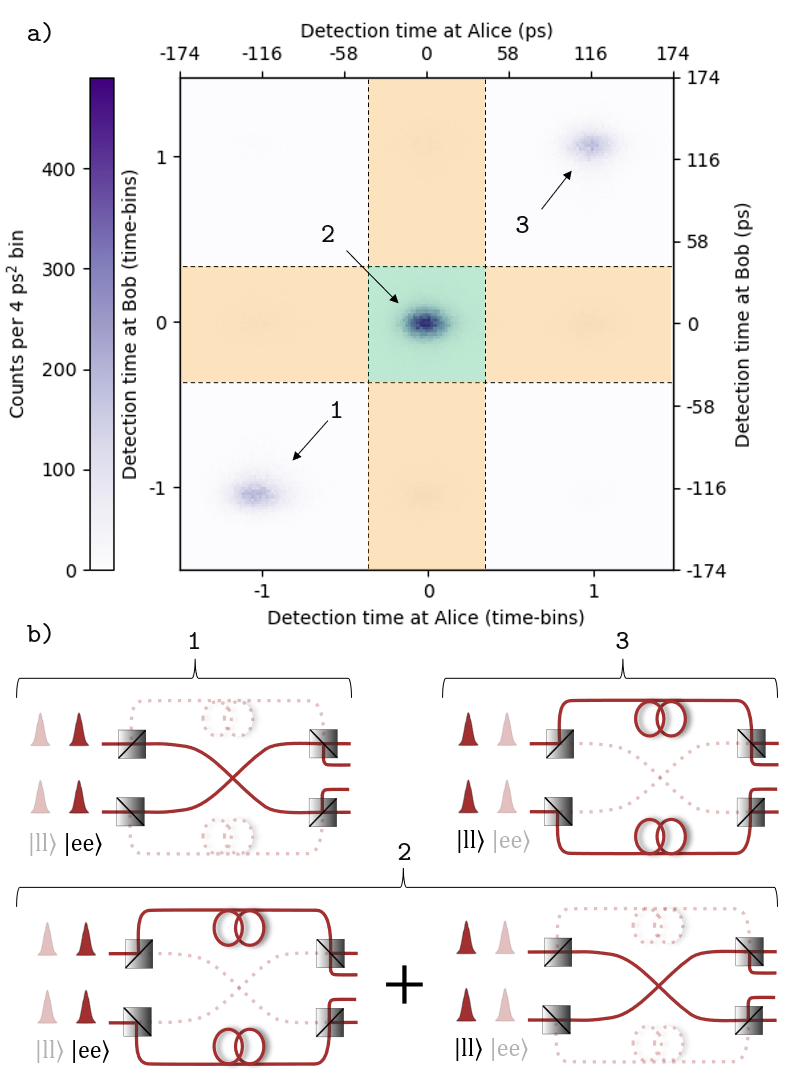}
  
  \caption{a) Joint histogram of detection time moduli at Alice and Bob using the hug interferometer in our PIC. The vertical and horizontal bands represent the post-selection windows of Alice and Bob. With a Franson interferometer the orange area contains on average twice as many detections as the ones in the green area which exhibit interference (see Fig.~\ref{fig:arrivals_distribution_scheme}), so a Bell violation is only possible by post-selecting the intersection of the two bands, which introduces a loophole. Instead, with the hug configuration the orange area ideally does not contain any counts so the union of the two bands, which is a valid local post-selection, can be taken without spoiling the violation. b) Case 1 (3) corresponds to pairs generated by the early (late) pump pulse taking the short (long) paths. Case 2 corresponds to the superposition where photons from the early pump pulse taking the long paths interfere with the photons from the late pump pulse taking the short paths.
  }
  \label{fig:joint_histogram}
\end{figure}

%%%%%%%%%%%%%%%%%%%%%%%%%%%%%%%%%%%%%%%%%%%%%%%%%%%%%%%%%%%%%%%%%%%%%%%%%%%%%%%%
%\vskip0.5cm
\begin{figure*}[t]
  \centering
  \includegraphics[width=1\linewidth,trim=0cm 1cm 0cm 0.5cm,clip]{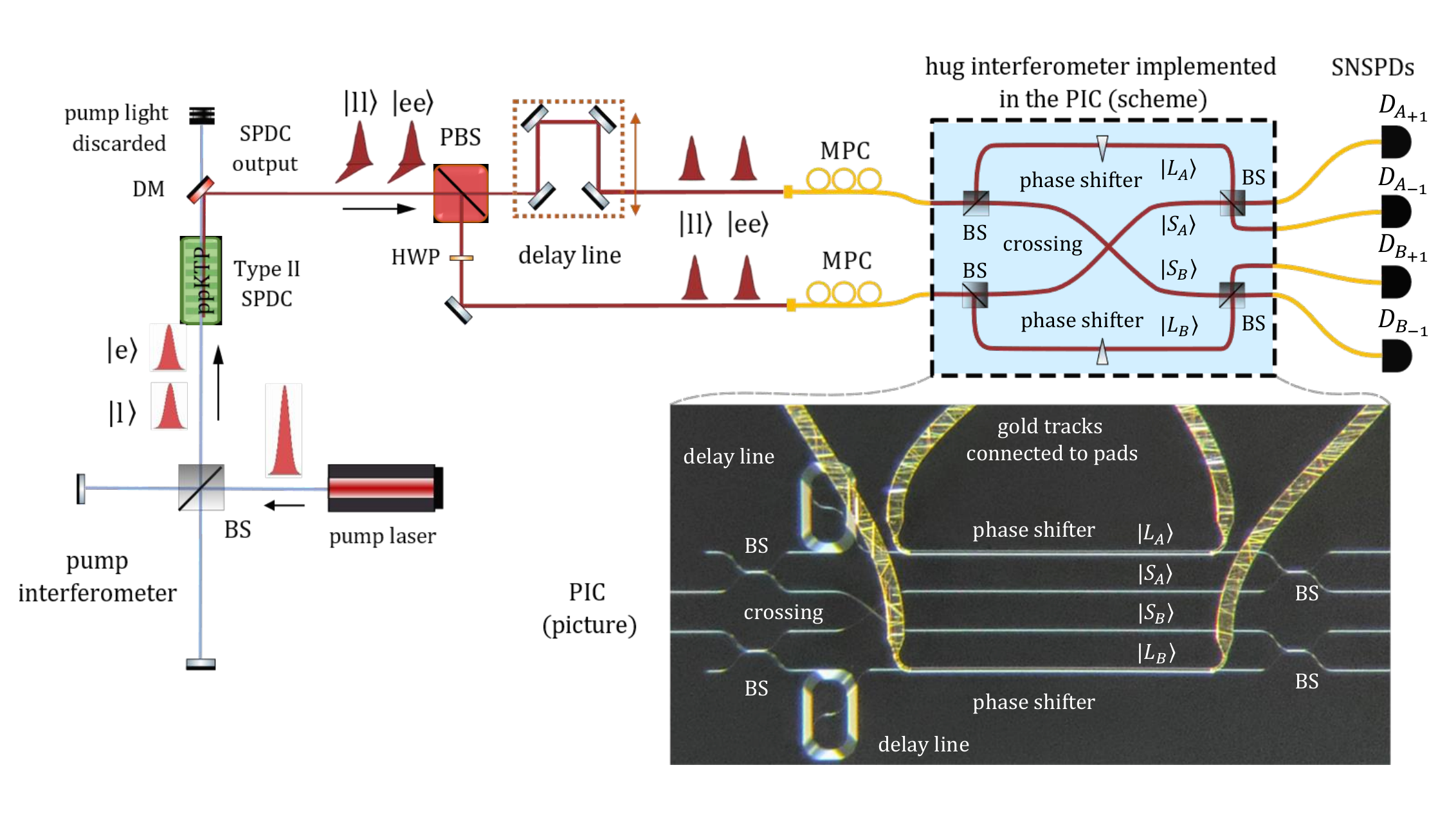}
  \caption{Overview scheme of the experimental setup used to generate time-bin entangled states and certify their entanglement. DM: dichroic mirror, PBS: polarizing beam-splitter, BS: beam-splitter, MPC: manual polarization controllers, PIC: photonic integrated chip, HWP: half-wave plate, SNSPDs: superconducting nanowire single-photon detectors, D: detector. }
  \label{fig:setup_scheme}
\end{figure*}

%%%%%%%%%%%%%%%%%%%%%%%%%%%%%%%%%%%%%%%%%%%%%%%%%%%%%%%%%%%%%%%%%%%%%%%%%%%%%%%%

In the hug configuration, the two measurement interferometers are connected such that detections at both Alice and Bob can only be due to the two down-converted photons propagating on paths of the same length: either both taking the short paths or both taking the long paths.
In the ET case, it is thus sufficient to discard events where two photons are received on the same side to be left with coincidence events that always result from the two-photon interference of these two indistinguishable processes.
In the TB case instead, as illustrated in Fig. \ref{fig:joint_histogram}, the two photons have three possible detection times with respect to the pump pulse emission even for coincidences. This leads in Bell tests to the same three-peaks histogram of arrival times that is obtained by using the Franson interferometer and only the photon pairs detected in the central TB display interference (case 2 of  Fig. \ref{fig:joint_histogram}). Coincidences in the first and last TB can instead be associated with a single known emission time and propagation path (cases 1 and 3 of Fig. \ref{fig:joint_histogram}).  Nonetheless, when looking at the time differences between detections at Alice and Bob we have a single peak as the generation time and the traveled optical path lengths are equal. This fact is illustrated in the upper pane of Fig. \ref{fig:joint_histogram} where most detections lie on the main diagonal, and will allow to close the post-selection loophole. Indeed, no communication between the two parties is needed for the post-selection which can be done locally without introducing any loophole. Alice and Bob locally post-select the detections forming the central peak of their arrival time histogram by defining an appropriate window and end up with the same set of events. In practice, it may happen that a detection happens on the central peak for one of the partners and on a lateral peak for the other. This can be caused by double pairs, failed separation of the signal and idler photons, and detection temporal jitter. As detailed in \ref{sec:AppendixA}, we can treat these events as losses by assigning a fixed outcome to the measurements as it is done to deal with the efficiency loophole and maintain the CHSH inequality for LHVM \cite{Larsson_2014}. These events will lower the violation but thanks to the geometry of the hug configuration they will remain very rare and mainly due to higher order emissions of the SPDC process.

%%%%%%%%%%%%%%%%%%%%%%%%%%%%%%%%%%%%%%%%%%%%%%%%%%%%%%%%%%%%%%%%%%%%%%%%%%%%%%%%

{\it Experimental setup.---}The experimental setup is shown in Fig.~\ref{fig:setup_scheme}. A mode-locked Ti:Sapphire laser is used to produce pump pulses of a few picoseconds at a rate of 76 MHz. The pulses are sent to a Michelson interferometer with an optical path-length imbalance between the two arms matching the imbalance between the short and long paths in the chip interferometers. Pulses are then coupled into a polarization-maintaining single-mode fiber and sent to a periodically-poled potassium titanyl phosphate (ppKTP) waveguide. The wavelength of the pump laser is tuned to 775 nm so that the two photons produced through the degenerate Spontaneous Parametric Down-Conversion (SPDC) process have a spectrum centered around 1550 nm, which enables the use of C-band components and efficient propagation in optical fiber-based telecommunication networks. The pump pulses are separated from the biphotons through a long-pass dichroic filter. Since type-II SPDC phase matching is used, signal and idler can be separated by a polarization beam-splitter (PBS) and then sent to the two inputs of the PIC. The PIC inputs and outputs are each butt-coupled via a fiber array using fiber alignment stages with micrometric precision. In contrast to what happens with a Franson interferometer, when using the hug configuration (with symmetric delays) it is essential to match the time of arrival of the two photons wavepackets at the input beam-splitters (BS) of the interferometer (see \ref{sec:AppendixB}). A free-space delay line, built using a micrometric stage, was needed before the PIC to compensate for this mismatch. 

As anticipated, the hug interferometer (with its phase modulators) is integrated into the photonic chip based on the Triplex silicon nitride (Si$_3$N$_4$) platform manufactured by LioniX \cite{Triplex}. This platform employs an asymmetric double-stripe cross-section structure for the waveguides surrounded by silicon dioxide (SiO$_2$), which is optimal for propagation due to the low losses. The high contrast in the refractive index also allows for low bending losses, which is very relevant to implement delays necessary for the construction of the hug structure. The schematic of the chip is also shown in Fig.~\ref{fig:setup_scheme}, where spot size converters are employed to couple the light from single-mode telecom fibers to match the mode of the double-stripe waveguides on the chip. Following the two input paths, two 50:50 bidirectional couplers, working as beamsplitters (BS), are employed to create four parallel paths. Two long delay lines, corresponding to 116 ps time delay, made from waveguide spirals are placed in two of the paths to create the long arms, which are then followed by two thermal phase shifters. These are controlled by the heat produced from an electrical current flowing through a thin gold wire deposited on top of the waveguide. The other two parallel arms (short), are crossed to create the hug configuration by connecting to the opposite party. Then the two upper and lower paths are connected to two 50:50 bidirectional couplers where the final joint projection is made. The four outputs are then led out of the chip through the use of spot-size converters, where an array of single-mode fibers are aligned at the chip's edge and reach superconducting nanowire single-photon detectors (SNSPDs) with on average: detection efficiency of $69\%$, RMS jitter of $8.3$ ps and dark-counts rate of 100 Hz. The overall insertion loss from the PIC is 8.8 dB, of which more than 6 dB are due to coupling.

Implementing the hug interferometer fully on a PIC allows for intrinsic phase stability and further shows the versatility of the hug interferometer as a characterization device by working for both TB, as shown here and for ET sources, as shown in previous works \cite{cuevas_long-distance_2013, lima_experimental_2010, carvacho_postselection-loophole-free_2015}.

%%%%%%%%%%%%%%%%%%%%%%%%%%%%%%%%%%%%%%%%%%%%%%%%%%%%%%%%%%%%%%%%%%%%%%%%%%%%%%%%
%\vskip0.5cm

\begin{figure}[t]
  \centering
  \includegraphics[width=\linewidth,trim=0cm 0cm 0cm 0cm,clip]{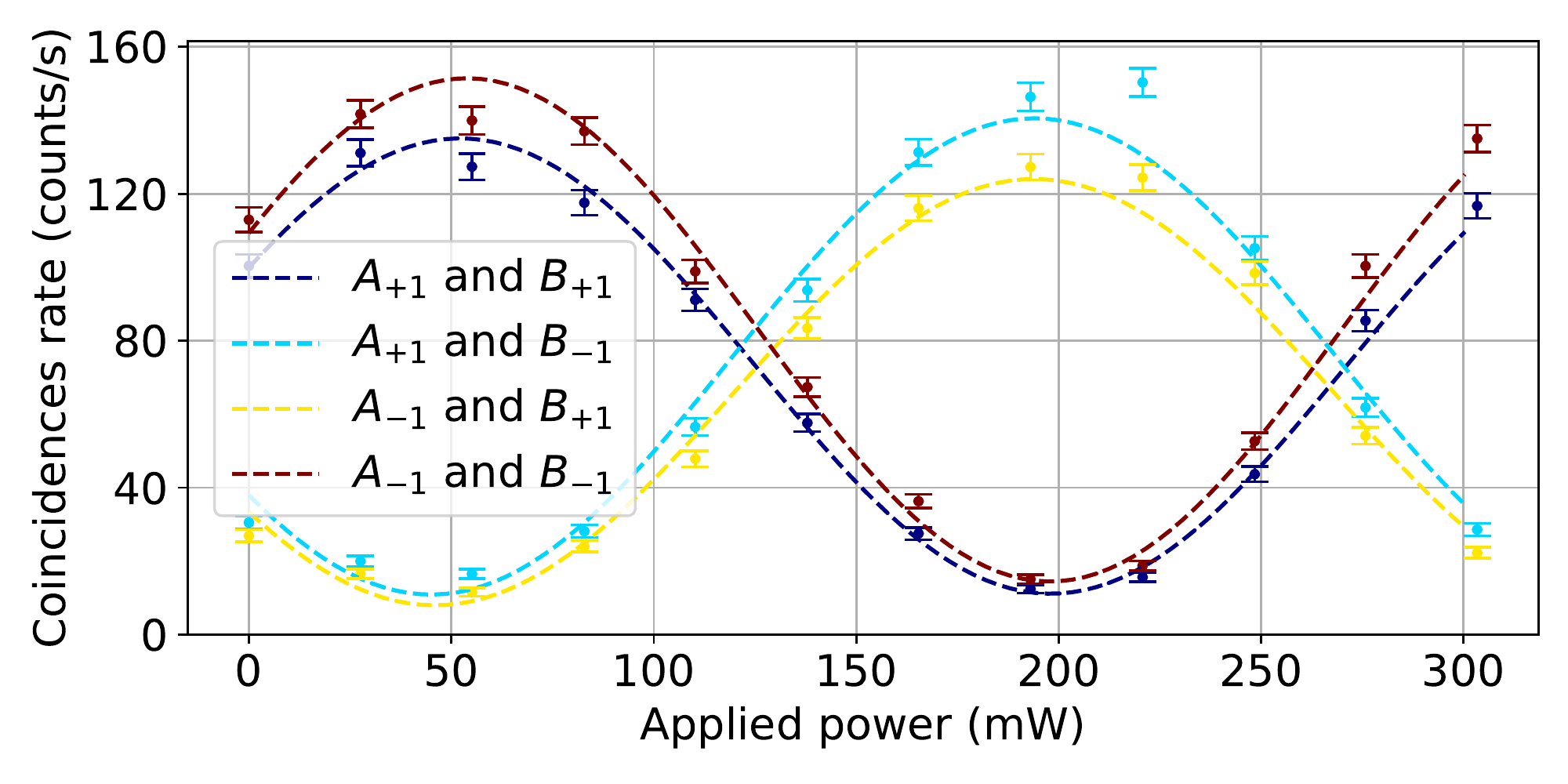}
  \caption{Modulation by interference of the coincidence rates in the central peak (100 ps window) versus power applied to Alice thermal phase-shifter. Measured data as dots with one standard deviation error bars and sinusoidal fit as dashed lines. 
  }
  \label{fig:phase_shifters_scan}
\end{figure}

{\it Results.---}The input state was tested by attempting a violation of the CHSH inequality. The measurements in the two bases at Alice and Bob were performed by tuning the phase differences between the long and the short arms of the interferometers to the appropriate values. To calibrate the thermal phase-shifters, a scan over voltages of the post-selected coincidence rates was performed, leading through two-photon interference to the characteristic sinusoidal modulation of the coincidence counts as shown in Fig.~\ref{fig:phase_shifters_scan}. 

%%%%%%%%%%%%%%%%%%%%%%%%%%%%%%%%%%%%%%%%%%%%%%%%%%%%%%%%%%%%%%%%%%%%%%%%%%%%%%%%

%%%%%%%%%%%%%%%%%%%%%%%%%%%%%%%%%%%%%%%%%%%%%%%%%%%%%%%%%%%%%%%%%%%%%%%%%%%%%%%%

To minimize the reduction of the Bell parameter due to the local post-selection procedure, it is essential to have a separation between the time-bins allowing for clear discrimination of the events belonging to the central peaks and to the lateral peaks. Assuming a Gaussian distribution of the detection time moduli for each time-bin, the delay of 116 ps between long and short arms in our PIC required us to lower the standard deviation of such Gaussians around 20 ps to have a negligible impact on the CHSH violation. The use of low-jitter SNSPD detectors (Single Quantum Eos) and time-taggers (Qutools QuTAG), along with a high precision synchronization to the fluctuations of the pump laser repetition rate allowed us to achieve this. First, we derived through frequency synthesis on an FPGA a signal at 10 MHz locked to the 76 MHz signal of the pump laser, which was used as clock for our time-tagger. Then, given that we could still observe a drift in the average detection time modulus for each time-bin, we implemented a real-time interpolation algorithm allowing us to correct for this drift and achieve arrival time distributions with standard deviations as low as $8.2\pm0.1$ ps as shown in Fig.~\ref{fig:joint_histogram} when detection rates are high enough to sample the drift.

%%%%%%%%%%%%%%%%%%%%%%%%%%%%%%%%%%%%%%%%%%%%%%%%%%%%%%%%%%%%%%%%%%%%%%%%%%%%%%%%

\begin{figure}[b]
  \centering
  \includegraphics[width=\linewidth]{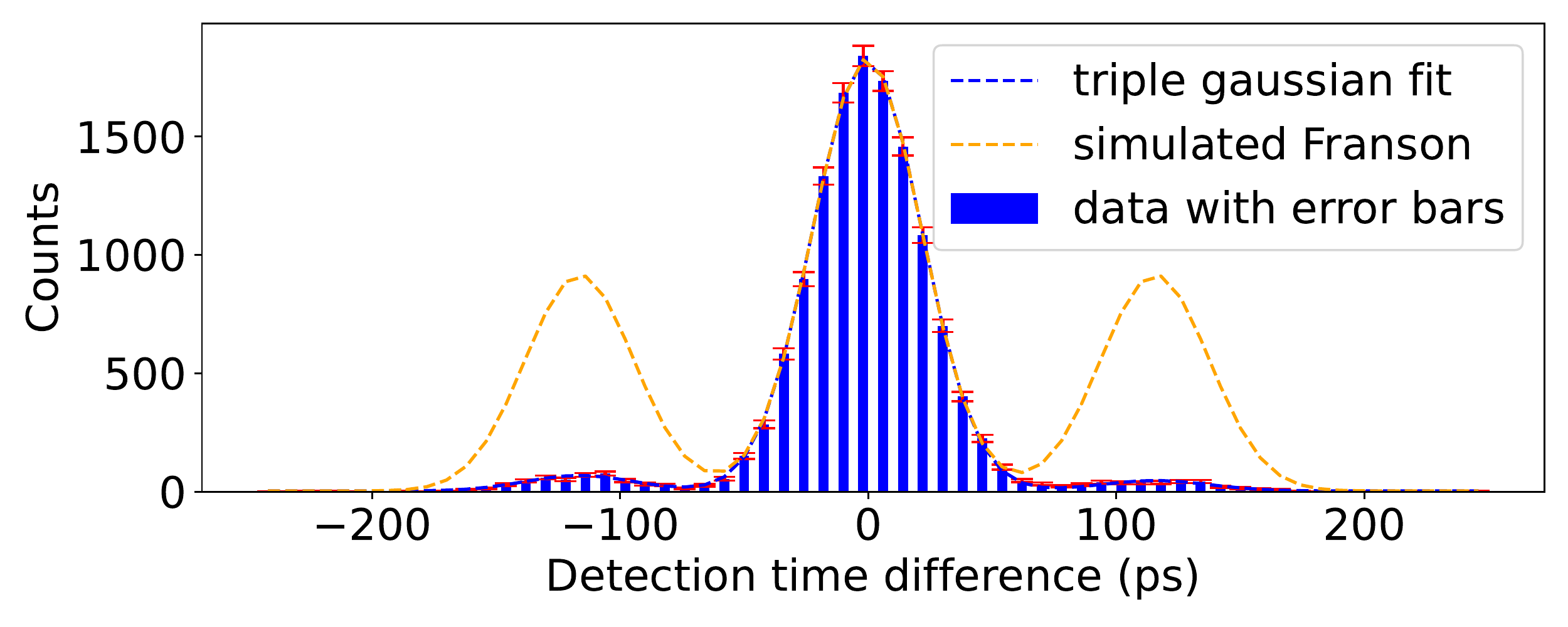}
  \caption{Histogram of detection time differences for arrivals at detectors $D_{A_{+1}}$ and $D_{B_{+1}}$ with bin-width of 8 ps. Data measured during the Bell test with one standard deviation error bars and best fit using as model the mixture of three Gaussian distributions with means shifted by one time-bin each. Note how the lateral peaks are reduced with respect to the simulated Franson scheme. From the fit, the lateral peaks jointly account for about $5.7\%$ of the events. The larger width of the Gaussian with respect to Fig.~\ref{fig:joint_histogram} is due to the lower biphoton generation rate used in the Bell test to reduce double pairs, which impacts the clock drift correction algorithm.}
  \label{fig:detection_time_difference}
\end{figure}

%%%%%%%%%%%%%%%%%%%%%%%%%%%%%%%%%%%%%%%%%%%%%%%%%%%%%%%%%%%%%%%%%%%%%%%%%%%%%%%%

The Bell test was performed by measuring the correlations in the four different basis combinations by maintaining the corresponding phase-shifters voltages in a predetermined sequence. A value of $S_{LP} = 2.42 \pm 0.05 $ was obtained which represents a violation of the CHSH inequality by more than 7 standard deviations. We identified several factors lowering the value of the violation. First, the visibility of the two-photon interference in the hug configuration depends on the indistinguishability between the signal and idler photons, which we assessed through Hong-Ou-Mandel interference to be $94.7\pm0.3\%$ for this Type II SPDC source. Second, the output beam-splitters of our interferometer were not perfectly balanced, we measured power-splitting ratios of $55.6 \pm 0.1 \%$ and $56.02 \pm 0.04 \% $, similar fabrication imperfections of Silicon Nitride directional couplers were also mentioned in \cite{Zhang:18}. The interferometric visibility measured in the central peak by scanning the pump interferometer was $89.2 \pm 0.7\%$ for around seven hundred coincidences per second. The experiment also suffered from thermal cross-talk between the phase-modulators, introducing a phase drift, which reduced the violation and limited to a dozen seconds the measurement time in each setting. Longer acquisitions led to violations of more than 10 standard deviations at the price of a reduced Bell parameter. This problem could be avoided by further distancing between the phase-shifters, the use of isolation trenches, or thermoelectric cooling. Finally, the local post-selection procedure also reduces the Bell parameter by about $0.06$ with respect to the non-local post-selection. Even if this might be considered a fair price for closing the post-selection loophole, we stress that this reduction is only due to the experimental parameters mentioned at the end of the section \textit{Genuine time-bin entanglement} which lead to non-null lateral peaks in the arrival-time differences histogram shown in Fig.~\ref{fig:detection_time_difference}, but could be improved.

%%%%%%%%%%%%%%%%%%%%%%%%%%%%%%%%%%%%%%%%%%%%%%%%%%%%%%%%%%%%%%%%%%%%%%%%%%%%%%%%

%\label{sec:Conclusion}
{\it Conclusions.---} We have presented the first implementation of a hug interferometer inside a photonic integrated circuit (PIC) and have shown its usefulness as a certification tool for both genuine (i.e., post-selection loophole-free) energy-time (ET) and time-bin (TB) entanglement. Using a PIC hug interferometer, we have provided the first violation of a Bell inequality using genuine TB entangled photons and the hug configuration, reporting a CHSH-Bell parameter of $2.42 \pm 0.05$. Demonstrating the use of a PIC to certify sources of genuine TB/ET entanglement is an important achievement in the field of quantum communication and a step forward toward a secure quantum communication platform. Future steps to improve the presented design could include mitigation of thermal cross-talk in the PIC and the use of tunable beam-splitters implemented through Mach-Zehnder interferometers \cite{Ma_2011} to increase the interferometric visibility.

%%%%%%%%%%%%%%%%%%%%%%%%%%%%%%%%%%%%%%%%%%%%%%%%%%%%%%%%%%%%%%%%%%%%%%%%%%%%%%%%

\begin{acknowledgments}
We would like to thank Single Quantum for kindly lending us their Eos SNSPDs whose low detection time jitter was essential to achieve the presented results and LioniX for manufacturing the integrated photonics chip.

%%%%%%%%%%%%%%%%%%%%%%%%%%%%%%%%%%%%%%%%%%%%%%%%%%%%%%%%%%%%%%%%%%%%%%%%%%%%%%%%

\textbf{Funding}: SECRET (SECuRe
quantum communication based on Energy-Time/time-bin entanglement): this project was funded within the QuantERA Programme that has received funding from the European Union’s Horizon 2020 research and innovation program under Grant Agreement No.\ 731473 and 101017733 and VR-2019-00392 (\href{http://dx.doi.org/10.13039/501100011033}{MCINN/AEI} Project No.\ PCI2019-111885-2). MIUR (Italian Minister for Education) under the initiative ``Departments of Excellence'' (Law 232/2016). A.A.\ and G.B.X.\ acknowledge the financial support of Ceniit Linköping University, the Swedish Research Council (VR 2017-0447) and the Knut and Alice Wallenberg Foundation through the Wallenberg Center for Quantum Technology (WACQT).

\end{acknowledgments}

%%%%%%%%%%%%%%%%%%%%%%%%%%%%%%%%%%%%%%%%%%%%%%%%%%%%%%%%%%%%%%%%%%%%%%%%%%%%%%%%

%%%%%%%%%% Merge with sappendix %%%%%%%%%%
\pagebreak
\widetext
\begin{center}
\textbf{\large Appendix}
\end{center}
%reset counters and prepend A to equation numbers in view of merging with paper

\setcounter{equation}{0}
\newcounter{reached_supplemental}
\setcounter{reached_supplemental}{0}
\counterwithin{figure}{reached_supplemental}
\stepcounter{reached_supplemental}
\setcounter{table}{0}
\setcounter{page}{1}
\makeatletter
\renewcommand{\theequation}{A\arabic{equation}}
\renewcommand{\thefigure}{A\arabic{figure}}
\renewcommand{\thesection}{A-\Roman{section}}

\section{\label{sec:AppendixA} Genuine time-bin entanglement using an interferometer in hug configuration}

In this section, we aim to provide the reader with a more detailed explanation of the use of the ``hug'' configuration interferometer to assess genuine time-bin entanglement. First, we will describe what state is obtained by feeding a time-bin entangled pair into an interferometer in the hug configuration and what is the resulting joint distribution of detection times of photons at the two analysis stations (that we will colloquially call Alice and Bob). Then we will tackle the issue of the post-selection loophole itself by describing a post-selection procedure adapted to the time-bin case which is completely local and doesn't open any loophole in the Bell test.

\subsection{\label{sec:PSLF_TB_1} Scheme description}

\begin{figure}[t]
  \centering
  \includegraphics[width=\linewidth,trim=0cm 0cm 0cm 0cm]{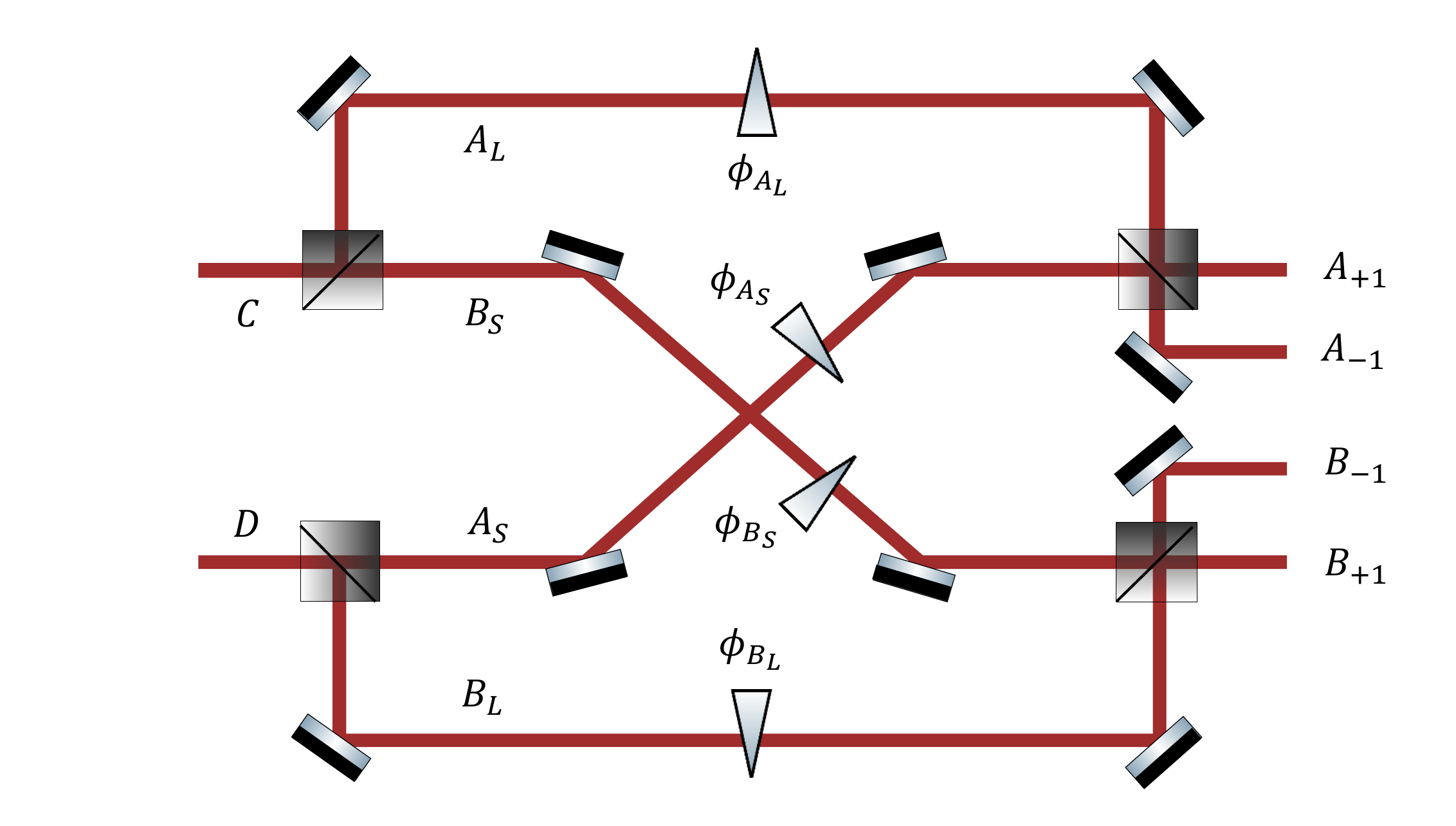}
  \caption{Optical scheme representing with equivalent bulk components the interferometer in hug configuration integrated into the PIC. Labels indicates the different spatial modes and phase-shifts ($\phi_{A_L}$, $\phi_{A_S}$, $\phi_{B_S}$, $\phi_{B_L}$).}
  \label{fig:hug_bulk_scheme}
\end{figure}

\begin{figure*}[htb]
  \centering
  \includegraphics[width=\linewidth,trim=0cm 0cm 0cm 0cm]{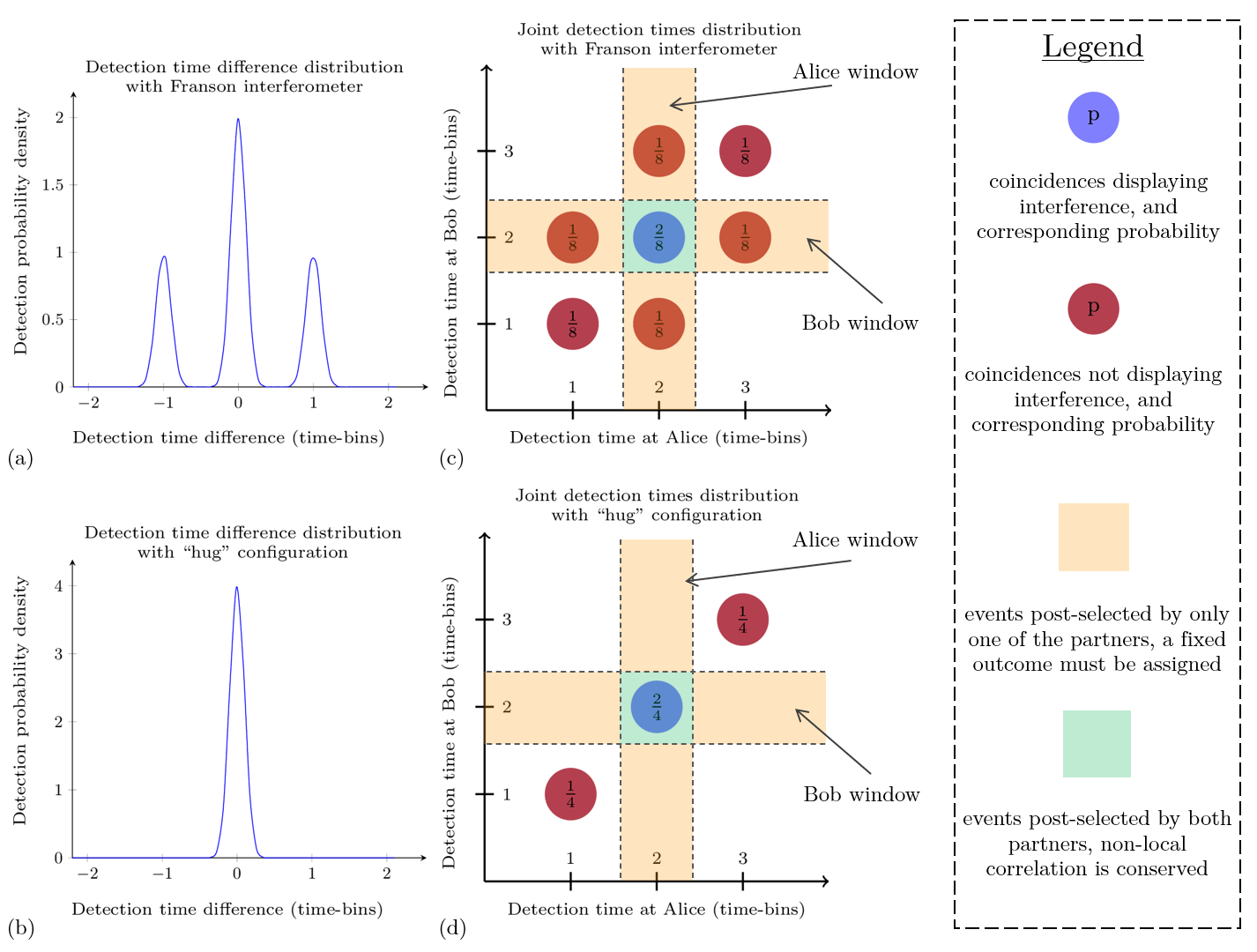}
  \caption{On the left, theoretical distributions of the detection time difference (between Alice and Bob detectors) showing three peaks with Franson interferometer (a) and a single one for the hug configuration (b). On the right, schemes showing the joint probability distribution of photon detection times at Alice and Bob measurement beam-splitters when using a Franson interferometer (c) and a hug interferometer (d) in the time-bin case. With the hug configuration, one photon is only detected at both Alice and Bob detectors if the two photons have traveled the same length, thus the time correlation of the input state is conserved. Hence Alice and Bob by locally post-selecting detections on time-bin 2 will select the same set of events, which are displaying full interferometric visibility. This is not possible for the Franson interferometer as a detection in time bin 2 at one partner can correspond to a detection in any of the three time-bins at the other, thus reducing the maximal visibility to 50\%.}
  \label{fig:arrivals_distribution_scheme}

\end{figure*}

Consider the hug interferometer represented in Fig.~\ref{fig:hug_bulk_scheme} and the labelled input modes ($C$, $D$) and output modes of Alice ($A_{+1}$, $A_{-1}$) and Bob ($B_{+1}$, $B_{-1}$). We can define the corresponding creation operators at time-bin $n$ as: $c^\dag_n$, $d^\dag_n$, $a^\dag_{+,n}$, $a^\dag_{-,n}$, $b^\dag_{+,n}$ and $b^\dag_{-,n}$. After the full measurement interferometers, the transformation is 
\begin{align}
c^\dag_n&\rightarrow \frac{1}{2}[e^{i\phi_{A_L}}(-a^\dag_{+,n+1}+ia^\dag_{-,n+1})+e^{i\phi_{B_S}}(b^\dag_{+,n}+ib^\dag_{-,n})],
\\
d^\dag_n&\rightarrow \frac{1}{2}[e^{i\phi_{B_L}}(-b^\dag_{+,n+1}+ib^\dag_{-,n+1})+e^{i\phi_{A_S}}(a^\dag_{+,n}+ia^\dag_{-,n})].
\end{align}
The input state consists of a pair of maximally entangled time-bin qubits. It can be produced by a source composed of a pulsed pump laser, an unbalanced interferometer, and a nonlinear device that generates photon pairs. We consider the interferometer unbalance to be inferior to half the pump repetition rate so that wavepackets generated by successive pulses will not interfere. The input state can be expressed as:
\begin{equation}
\ket{\Phi_{in}} \triangleq \frac{1}{\sqrt{2}}(c^\dag_1d^\dag_1+e^{i\phi_p}c^\dag_2d^\dag_2)\ket0,
\end{equation}
and it is transformed by the measurement interferometers into
\begin{equation}
\begin{aligned}
\ket{\Phi_{in}}\rightarrow&
 \frac{1}{4\sqrt2}[e^{i\phi_{A_L}}(ia^\dag_{-,2}-a^\dag_{+,2})+e^{i\phi_{B_S}}(b^\dag_{+,1}+ib^\dag_{-,1})]
 [e^{i\phi_{B_L}}(ib^\dag_{-,2}-b^\dag_{+,2})+e^{i\phi_{A_S}}(a^\dag_{+,1}+ia^\dag_{-,1})]\ket0
 +
 \\
& + \frac{e^{i\phi_p}}{4\sqrt2}[e^{\phi_{A_L}}(ia^\dag_{-,3}-a^\dag_{+,3})+e^{\phi_{B_S}}(b^\dag_{+,2}+ib^\dag_{-,2})]
 [e^{i\phi_{B_L}}(ib^\dag_{-,3}-b^\dag_{+,3})+e^{i\phi_{A_S}}(a^\dag_{+,2}+ia^\dag_{-,2})]\ket0.
\end{aligned}
\end{equation}
From the above equation, we can note that two photons can be detected in different time-bins if and only if they are both detected on the same measurement station, either at Alice or at Bob. This will happen with probability $\frac{1}{2}$. Let's now focus on the subspace where one photon arrives at Alice and one at Bob, in any time-bin
\begin{equation}
\begin{aligned}
&
 \frac{e^{i(\phi_{A_L}+\phi_{B_L})}}{4\sqrt2}[
 e^{i(2\Delta\phi-\phi_p)}(b^\dag_{+,1}+ib^\dag_{-,1})(a^\dag_{+,1}+ia^\dag_{-,1})
 +
 e^{i\phi_p}(ia^\dag_{-,3}-a^\dag_{+,3})(ib^\dag_{-,3}-b^\dag_{+,3})
 ]\ket0
 \\
&+\frac{e^{i(\phi_{A_L}+\phi_{B_L}+\Delta\phi)}}{2\sqrt2}[
\cos\left(\Delta\phi\right)(a^\dag_{+,2}b^\dag_{+,2}-a^\dag_{-,2}b^\dag_{-,2})
-\sin\left(\Delta\phi\right)(a^\dag_{-,2}b^\dag_{+,2}+a^\dag_{+,2}b^\dag_{-,2})
]\ket0,
\end{aligned}
\end{equation}
where
\begin{equation}
    \Delta\phi\triangleq\frac12(\phi_p+\phi_{A_S}+\phi_{B_S}-\phi_{A_L}-\phi_{B_L}).
    \label{eqsup:deltaPhi}
\end{equation}
 Upon considering only the central time-bin on either partner (i.e. $a_{\pm,2}^\dag$ and $b_{\pm,2}^\dag$), the relevant term becomes
\begin{equation}
[
\cos\left(\Delta\phi\right)(a^\dag_{+,2}b^\dag_{+,2}-a^\dag_{-,2}b^\dag_{-,2})
-\sin\left(\Delta\phi\right)(a^\dag_{-,2}b^\dag_{+,2}+b^\dag_{-,2}a^\dag_{+,2})
]\ket0
\end{equation}
and displays full interferometric visibility. Conditioned on detection at Alice and Bob, this happens with probability $\frac{1}{2}$. Alternatively, both photons will be detected in time-bin $1$ or $3$ with equal probability of $\frac{1}{4}$. This distribution of detection times is represented in Fig.~\ref{fig:arrivals_distribution_scheme}, and it is compared to what one obtains using a Franson interferometer with the same input state.

\subsection{\label{sec:PSLF_TB_2} Local post-selection procedure}

The post-selection loophole deriving from the use of Franson's interferometer to assess time-bin entanglement is similar to the detection loophole \cite{Larsson_2014}. Indeed, using Franson's interferometer a violation of Bell inequalities is only observed if we estimate correlations from a subset of all the prepared pairs: the ones which are detected at both Alice and Bob in the central time-bin (which corresponds to only $\frac{1}{4}$ of all detections). Other detections lead indeed to uncorrelated outcomes. Without additional assumptions no violation of Bell inequalities can actually be obtained with Franson's interferometer when considering all events \cite{jogenfors_2014}. We stress the fact that to post-select the events displaying interference, information is needed from the two parties. Indeed, a detection in the central time-bin on one side can correspond to a detection in any time-bin on the other side. As we have seen, with the hug configuration when a photon is detected in the central time-bin by one partner it is also detected in the central time-bin by the second partner. This is the crucial difference with respect to the Franson case as can be seen in Fig.~\ref{fig:arrivals_distribution_scheme}. 

In fact, this allows to perform a local post-selection of events. Alice and Bob can, independently, post-select the detections in the central time-bin and will end up with the same set of events. Obtaining an equal set of events through local post-selection guarantees that the set of post-selected coincidences is independent of the measurement settings. Suppose that the measurement bases are randomly switched fast enough to guarantee space-like separation of the choices. In that way, the post-selection at Alice cannot depend on the measurement choice of Bob and vice-versa. The post-selection could still depend on the local measurement settings, but if it did in some cases, one detection might be selected by Alice and independently discarded by Bob or vice-versa, as the selections would reduce to independent Bernoulli trials whose success probability might depend on a common hidden variable but which are otherwise independent of each-other.

In practice, it may happen that a detection happens on the central peak for one of the partners and on a lateral peak for the other. This can mainly be caused by double pairs, failed separation of the signal and idler photons, and, finally, detection temporal jitter. Considering this, the independence from local phase settings of the post-selection can no longer be guaranteed, but the post-selection can be treated as losses in the detection loophole by assigning a fixed outcome in cases where one user discarded an event while the other post-selected it. Local post-selection is indeed valid as long as the union of events post-selected at each measurement station is considered (and not only their intersection, which would lead again to the post-selection loophole). 

The complete post-selection procedure to obtain a valid violation is thus the following:
\begin{enumerate}
\item Alice and Bob tell each other for which pump pulses they detected a photon in any of their detectors and in any time-bin. Under the fair sampling hypothesis, the subset of events where both photons were detected is representative of the totality of events.
\item The 50\% of events where both photons are detected by the same partner are rightfully discarded as the selected events are independent of remote and local phase settings (this can be verified by switching randomly the phase settings at a high enough rate) \cite{cabello_proposed_2009}.
\item Alice and Bob locally post-select the detections forming the central peak of their detection time histogram by defining an appropriate window. They communicate to each other which events they have selected. If only one of the partners selected an event, the other one considers the twin photon as lost (even if actually it was received but not in the central peak) and assigns a fixed value to the outcome of the measure for that run, as it is done to deal with the efficiency loophole. 
\end{enumerate}
With a Franson interferometer, step 3 would bring the maximum value of the Bell parameter to $\frac{2\sqrt{2}}{3}<2$, as a central detection at one side corresponds with equal probability to a detection at the other side in the central time-bin or in the lateral ones and so only $\frac{1}{3}$ of events would display correlations. With the hug configuration instead, only non-idealities in the input state or in the experimental setup lead to events post-selected solely at one side, so if the two-photon interference visibility is sufficient we will generally still be able to obtain a violation.

\section{\label{sec:AppendixB}Indistinguishability condition for interference in the hug configuration}

In this section, we aim to highlight a feature of the hug configuration that was not explicitly mentioned in previous publications \cite{
rossi_generation_2008, cabello_proposed_2009,lima_experimental_2010,cuevas_long-distance_2013,carvacho_postselection-loophole-free_2015}  but might be of interest to the experimentalist. As we will show, the visibility of two-photon interference in the hug configuration depends on the indistinguishability of the two input photons, which is not the case with Franson interferometer. Of course, this follows from the fact that the interfering processes are the one where the signal photon goes to Alice and the idler photon goes to Bob and the one where the signal goes to Bob and the idler to Alice. If, in the detection process, some information about which of the two photons is detected on each side can be obtained (even in principle), then, which‐path information is obtained at the same time, leading to a reduction of the interference visibility. In our derivation in Sec.~\ref{sec:PSLF_TB_1}, we considered identical wavepackets at each input. But what if the two wavepackets have distinguishable temporal profiles despite belonging to the same time-bin?

The wavepacket of the photon in $C$ at time-bin $n$ is rewritten as $c^\dag_n\ket0$ and the wavepacket of the photon in $D$ is rewritten as $\widetilde d^\dag_n\ket0$, where the tilde will mark the difference between the two input photons.
Now, after the full measurement interferometers, the transformation is 
\begin{align}
c^\dag_n&\rightarrow \frac{1}{2}[e^{i\phi_{A_L}}(-a^\dag_{+,n+1}+ia^\dag_{-,n+1})+e^{i\phi_{B_S}}(b^\dag_{+,n}+ib^\dag_{-,n})],
\\
\widetilde d^\dag_n&\rightarrow \frac{1}{2}[e^{i\phi_{B_L}}(-\widetilde b^\dag_{+,n+1}+i\widetilde b^\dag_{-,n+1})+e^{i\phi_{A_S}}(\widetilde a^\dag_{+,n}+i\widetilde a^\dag_{-,n})].
\end{align}
With respect to the previous case, now we have different output modes creation operators (with and without the tilde), corresponding to slightly distinguishable wavepackets.
The input state is now written as 
\begin{equation}
\ket{\Phi_{in}} \triangleq \frac{1}{\sqrt{2}}(c^\dag_1\widetilde d^\dag_1+e^{i\phi_p}c^\dag_2\widetilde d^\dag_2)\ket0
\end{equation}
and transformed by the measurement interferometers into
\begin{equation}
\begin{aligned}
\ket{\Phi_{in}}\rightarrow&
 \frac{1}{4\sqrt2}[e^{i\phi_{A_L}}(ia^\dag_{-,2}-a^\dag_{+,2})+e^{i\phi_{B_S}}(b^\dag_{+,1}+ib^\dag_{-,1})][e^{i\phi_{B_L}}(i\widetilde b^\dag_{-,2}-\widetilde b^\dag_{+,2})+e^{i\phi_{A_S}}(\widetilde a^\dag_{+,1}+i\widetilde a^\dag_{-,1})]\ket0
 +
 \\
& + \frac{e^{i\phi_p}}{4\sqrt2}[e^{i\phi_{A_L}}(ia^\dag_{-,3}-a^\dag_{+,3})+e^{i\phi_{B_S}}(b^\dag_{+,2}+ib^\dag_{-,2})]
 [e^{i\phi_{B_L}}(i\widetilde b^\dag_{-,3}-\widetilde b^\dag_{+,3})+e^{i\phi_{A_S}}(\widetilde a^\dag_{+,2}+i\widetilde a^\dag_{-,2})]\ket0 .
\end{aligned}
\end{equation}
If we focus on the contributions in the time-bin 2, where one photon arrives at Alice and one at Bob, then the relevant terms are written as
\begin{equation}
\begin{aligned}
\ket{\chi_{2,2}}=&
 \frac{e^{i(\phi_{A_L}+\phi_{B_L})}}{4\sqrt2}
 \left[
 (a^\dag_{+,2}-ia^\dag_{-,2})
 (\widetilde b^\dag_{+,2}-i\widetilde b^\dag_{-,2})
 +
 e^{2i\Delta\phi}
(\widetilde a^\dag_{+,2}+i\widetilde a^\dag_{-,2})
(b^\dag_{+,2}+i b^\dag_{-,2})
 \right]\ket0,
\end{aligned}
\end{equation}
using again $\Delta\phi$ as defined in Eq.~\eqref{eqsup:deltaPhi}.
The probability of detecting at time-bin $2$ a photon in $A_\mu$ and one in $B_\nu$ with $\mu,\nu\in\left\{-1,+1\right\}$ is
\begin{equation}
\begin{split}
   p_{A_\mu,B_\nu}^{(2,2)}(\Delta\phi) &= \frac{1}{32}\left| 
   (a^\dag_{\mu,2}\widetilde b^\dag_{\nu,2}+e^{2i\Delta\phi_{\mu\nu}}
   \widetilde a^\dag_{\mu,2}b^\dag_{\nu,2})\ket0
   \right|^2
    = \frac{1}{16}[
   1+ |\gamma|^2 \cos\left(2\Delta\phi_{\mu\nu}\right)],
\end{split}
\end{equation}
where $\Delta\phi_{\mu\nu} \triangleq \Delta\phi +
\frac\pi2(\mu+\nu)$ and $\gamma$ is the overlap between the two inputs modes, namely,
\begin{equation}
\gamma=\bra{0} a_{\mu,2} \widetilde a^\dag_{\mu,2} \ket{0} = \bra{0} b_{\nu,2} \widetilde b^\dag_{\nu,2}\ket{0},
\end{equation}
whose squared modulus, as shown, represents the interferometric visibility in the central time-bins
\begin{equation}
    V_{A,B}^{(2,2)} = \frac{\frac{1}{16}\left(1 + |\gamma|^2\right)-\frac{1}{16}\left(1 - |\gamma|^2 \right)}{\frac{1}{16}\left(1 + |\gamma|^2 \right)+\frac{1}{16}\left(1 - |\gamma|^2 \right)} = |\gamma|^2.
\end{equation}

\begin{figure}[h]
  \centering
  \includegraphics[width=\linewidth,trim=0cm 0cm 0cm 0cm]{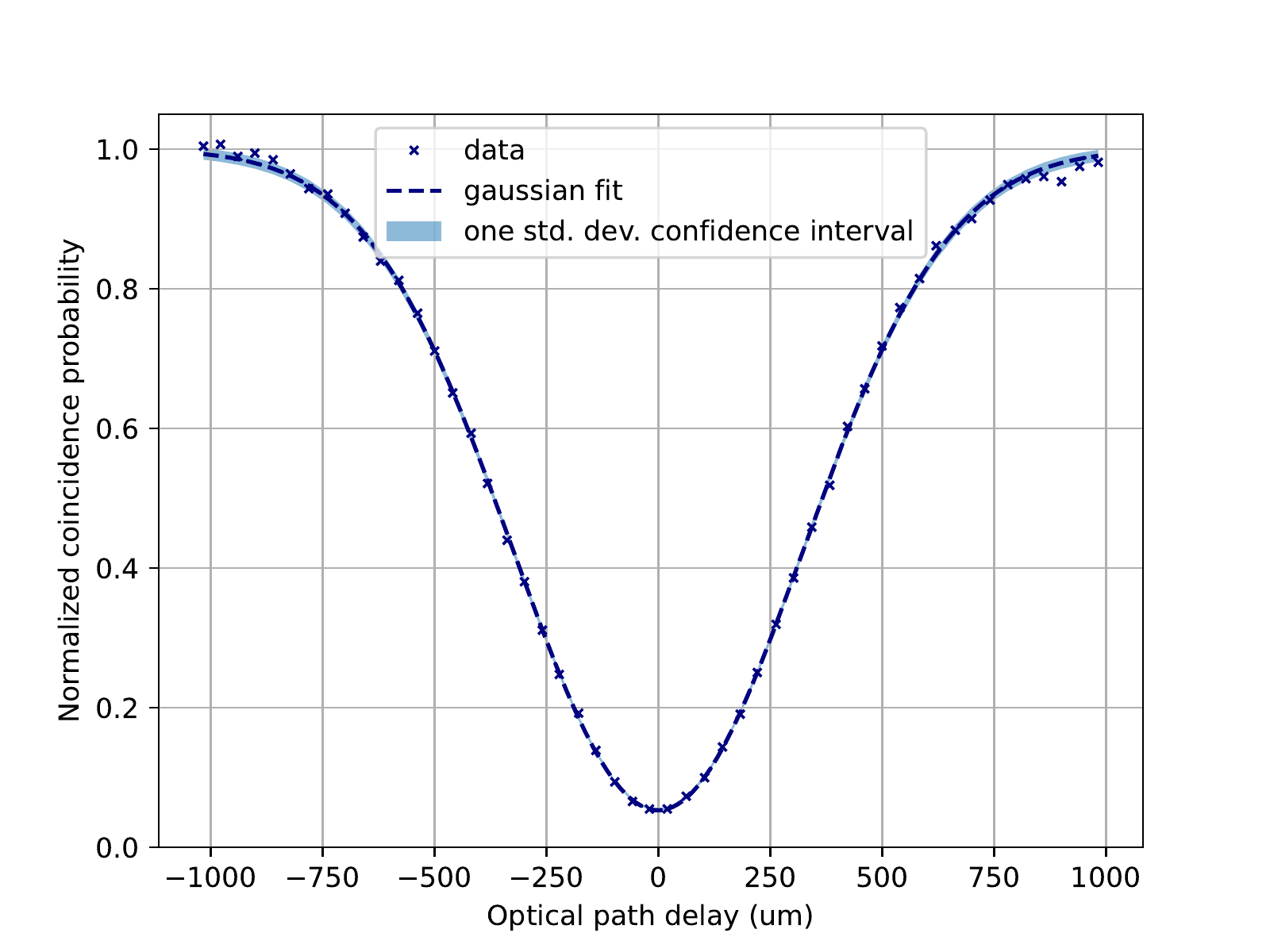}
  \caption{Hong-Ou-Mandel dip shown as a variation of the coincidence detection probability (normalized to 1) with the relative delay between the signal and idler photons impinging on the two inputs of a 50:50 beam-splitter. A coincidence window of $100$ ps was used. The area shaded in blue represents a confidence interval of one standard deviation assuming Poissonian counts.}
  \label{fig:HOM_dip}
\end{figure}

Note that this feature is not a consequence of using a time-bin state as input, the same applies to the energy-time case. This condition on the input state is analogous to the indistinguishability condition required to observe high contrast Hong-Ou-Mandel (HOM) interference \cite{ou_multi-photon_2007}. HOM interference can thus be used to characterize the indistinguishability of the two input photons and obtain an upper bound on the two-photon interference visibility obtained with any interferometer in hug configuration. We tested our SPDC source and obtained a visibility of $94.7\pm0.5\%$, the HOM dip is shown in Fig.~\ref{fig:HOM_dip}. We believe that spectral differences between signal and idler in type-II SPDC are responsible for this distinguishability as other degrees of freedom could be finely tuned.

\begin{figure}[h]
  \centering
  \includegraphics[width=\linewidth,trim=0cm 0cm 0cm 0cm]{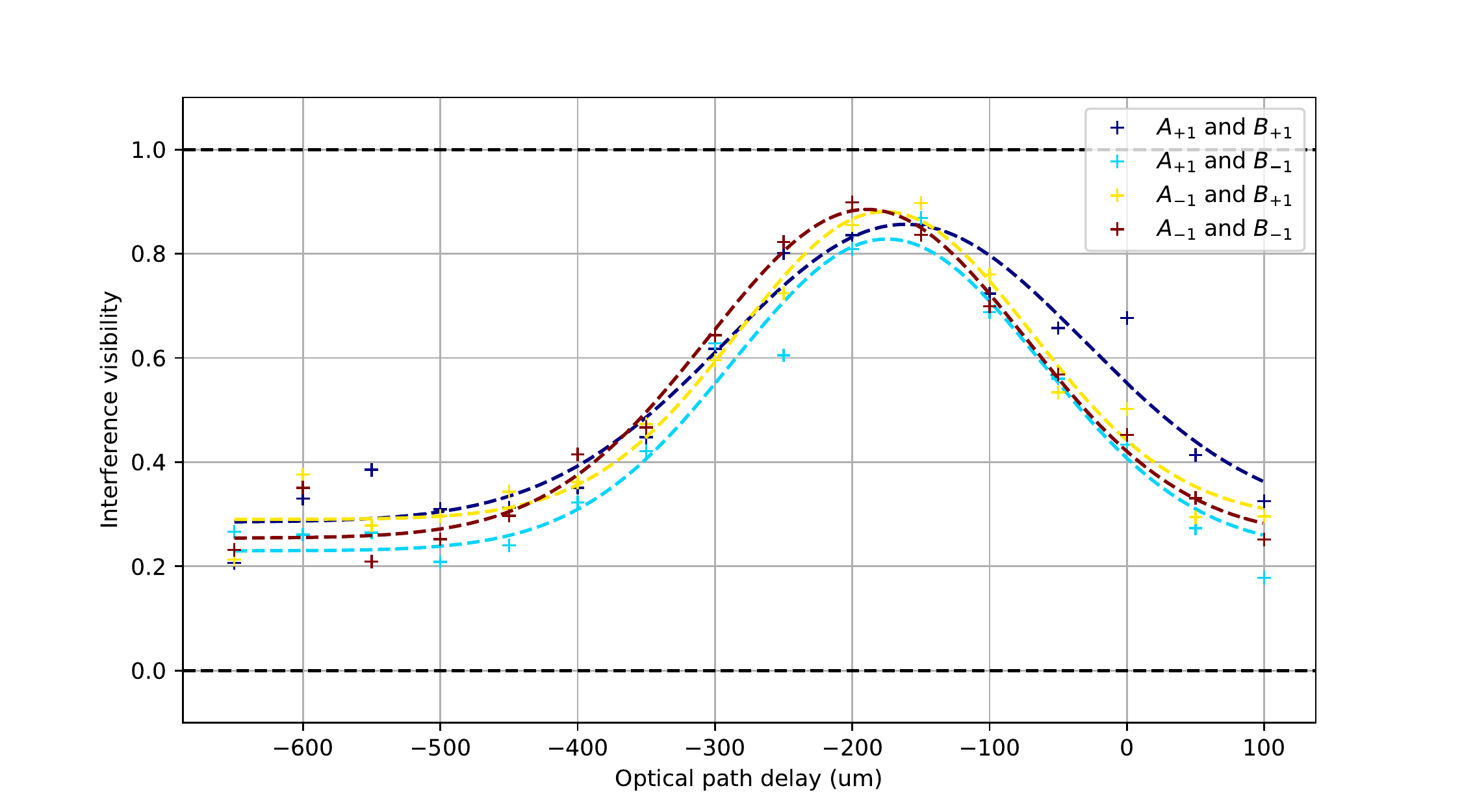}
  \caption{Scan of the linear stage controlling the delay between signal and idler before the inputs of the interferometer to find the position giving maximal visibility. At each position measures with different phase settings were taken to have an estimate of the visibility from the variation of the rates of coincidences between the different outputs. Best Gaussian fits are displayed as dashed lines.  Note that the estimated visibility doesn't drop to zero far away from the maximum due to statistical fluctuations in the coincidence counts.}
  \label{fig:signal_idler_delay}
\end{figure}

As indistinguishability in all degrees of freedom is required, it is essential to precisely match the arrival time of the signal and idler photons at the input beam-splitters of the hug interferometer. In our setup, the polarization dispersion in polarization-maintaining fibers was sufficient to spoil the interference and we thus had to insert in the setup an optical delay line to compensate for this delay. A scan of the delay to find the position of maximal interference is shown in Fig.~\ref{fig:signal_idler_delay}.

As we have seen the hug configuration requires the two photons to enter synchronously the input beam-splitters, which is not the case for Franson interferometer. This can be seen as a disadvantage but there is actually an upside to this feature. Consider the interferometer in hug configuration of Fig.~\ref{fig:hug_bulk_scheme}, the path differences between long and short arms for Alice and Bob should be equal in the ideal case \cite{cabello_proposed_2009}, that is,
\begin{equation}
      l_{A_L}-l_{A_S} = l_{B_L}-l_{B_S},
\end{equation}
where $l_{X}$ is the optical path-length of path $X$. Nonetheless, if the manufacturing process is inaccurate and the two differences are not equal it is still possible to obtain the maximal visibility interference in the hug configuration by tuning the delay between signal and idler photons before the interferometer. Let's first consider the energy-time case. As the generation time is unpredictable, the only temporal condition for indistinguishability of the short-short and the long-long processes is to have the difference between the detection time at Alice ($t_{A_a}$) and the one at Bob ($t_{B_b}$) to be equal for the two processes. Note that any delay after the output beam-splitters at Alice and Bob is irrelevant, so it can be set to zero without loss of generality. Consider $c$ to be the speed of light in vacuum and an arbitrary generation time $t_G$ for the biphoton, for the short-short case the difference is
\begin{equation}
      \tau_{ab}^{SS}= t_{A_a}^{SS} - t_{B_b}^{SS} = \left(t_G +   \frac{l_{D} + l_{A_S}}{c}\right) - \left(t_G + \frac{l_{C} + l_{B_S}}{c}\right) = \frac{1}{c} \left( l_{D} + l_{A_S} - l_{C} - l_{B_S} \right),
\end{equation}
while for the long-long case, we have
\begin{equation}
      \tau_{ab}^{LL}= t_{A_a}^{LL} - t_{B_b}^{LL} = \left(t_G +   \frac{l_{C} + l_{A_L}}{c}\right) - \left(t_G + \frac{l_{D} + l_{B_L}}{c}\right) = \frac{1}{c} \left( l_{C} + l_{A_L} - l_{D} - l_{B_L} \right).
\end{equation}
Therefore, the temporal indistinguishability condition $ \tau_{ab}^{SS} = \tau_{ab}^{LL}$ can be rewritten as
\begin{equation}
\begin{split}
    \frac{1}{c} \left( l_{D} + l_{A_S} - l_{C} - l_{B_S} \right) &= \frac{1}{c} \left( l_{C} + l_{A_L} - l_{D} - l_{B_L} \right) \\
    \iff  \left( l_{A_L} -l_{A_S}\right)  - \left( l_{B_L} - l_{B_S} \right) &= 2\left(l_{D}-l_{C}\right).\label{eqn:S21}
\end{split}
\end{equation}
From this last equation, it is clear that the delay between signal and idler photons before the interferometer can be used to compensate for mismatches in the delays inside the interferometer, which is not possible with a Franson interferometer and could be of particular interest when the interferometer is built with integrated optics.
In the time-bin case the possible generation times are known and to have interference we need to verify
\begin{equation}
    t_{A_a}^{LL} - t_{A_a}^{SS} = \Delta T = t_{B_b}^{LL} - t_{B_b}^{SS},
\end{equation}
where $\Delta T$ is the time difference between the time-bins which is determined by the delay in the pump interferometer. When Eq.~\eqref{eqn:S21} is satisfied, this leads to
\begin{equation}
\begin{split}
         \Delta T &= t_{A_a}^{LL} - t_{A_a}^{SS}   \\
          &=  \left(t_G +   \frac{l_{C} + l_{A_L}}{c}\right)   - \left(t_G +   \frac{l_{D} + l_{A_S}}{c}\right) \\
          &=  \frac{1}{c}\left(l_{C} + l_{A_L} -  l_{D} - l_{A_S}\right) \\
          &=  \frac{1}{c}\left( l_{A_L}- l_{A_S} - \frac{\left( l_{A_L} -l_{A_S}\right)  - \left( l_{B_L} - l_{B_S} \right)}{2} \right) \\
         &=  \frac{\left( l_{A_L} -l_{A_S}\right)  + \left( l_{B_L} - l_{B_S} \right)}{2c},
\end{split}
\end{equation}
thus the compensation is again possible provided the pump interferometer is set to have an unbalance which is the average of the two unbalances in the hug interferometer.

\section{\label{sec:AppendixC} Synchronization of the time-tagging system with the pump laser rate fluctuations}

\begin{figure}[ht]
  \centering
  \includegraphics[width=\linewidth,trim=0cm 0cm 0cm 0cm]{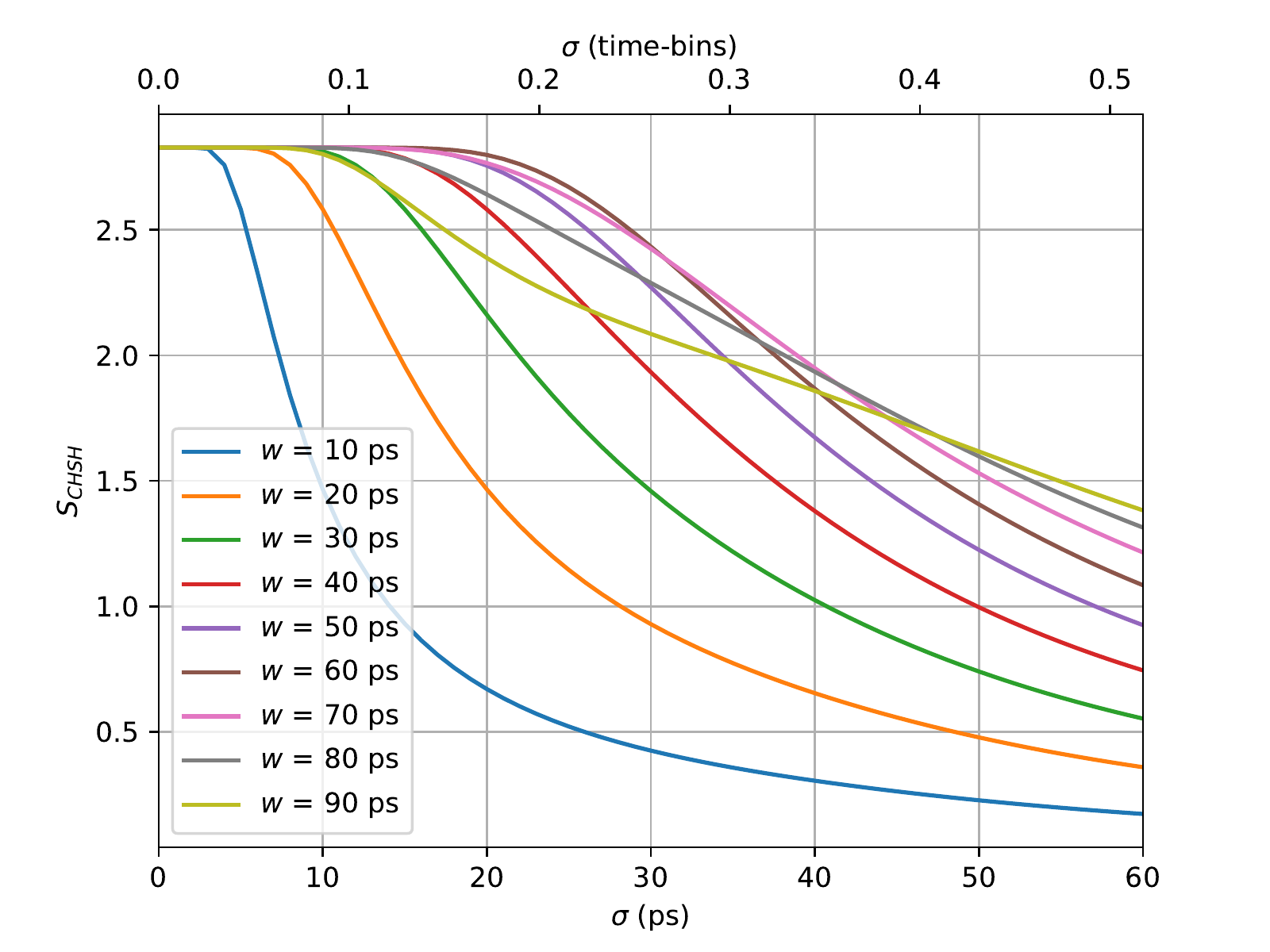}
  \caption{Theoretical maximum CHSH Bell parameter attainable using an interferometer in hug configuration in the time-bin case for different local post-selection window half-widths w. $S_{CHSH}$ is plotted against the RMS width $\sigma$ of detection time distributions of each time-bin (assuming independent Gaussian distributions with means shifted of $\Delta T = 116$ ps). The reduction of the Bell parameter is only due to incorrect discrimination of photons belonging to the different time-bins. Always considering the best window size, we can still see an important drop for $\sigma>20$ ps and after $\sigma>40$ ps no violation is possible at all (the classical limit of 2 is reached for $\frac{\sigma}{\Delta T}\approx0.335$).}
  \label{fig:CHSH_violation_vs_RMS_width}
\end{figure}

When using the hug configuration with a time-bin state a local post-selection of the detections in the central time-bin is necessary as detailed in \ref{sec:AppendixA}. If the ratio between the standard deviation of the detection time distribution and the time-bin separation is too low, detections in different time-bins cannot be clearly discriminated. This entails a reduction of the violation of the Bell inequality in two ways. First, we might post-select a pair of photons that actually belonged to the lateral time-bins and thus displays no correlation in the outcomes. Second, we might have cases where one photon is post-selected by one of the parties but not by the other because it was detected outside of its coincidence window. In that case, a fixed outcome (independent of the phase settings) has to be assigned, so the violation is also lowered. Given that the delay line in our PIC was only $2$ cm long, corresponding to about $\Delta T = 116$ ps of delay, it was essential for us to optimize the precision of the detection time measurements. In Fig.~\ref{fig:CHSH_violation_vs_RMS_width}, a simulation of the maximal Bell parameter attainable with local post-selection with respect to the standard deviation of detection times of photons belonging to a given time-bin is displayed.

\begin{figure}[h]
  \centering
  \includegraphics[width=\linewidth,trim=0cm 0cm 0cm 0cm,clip]{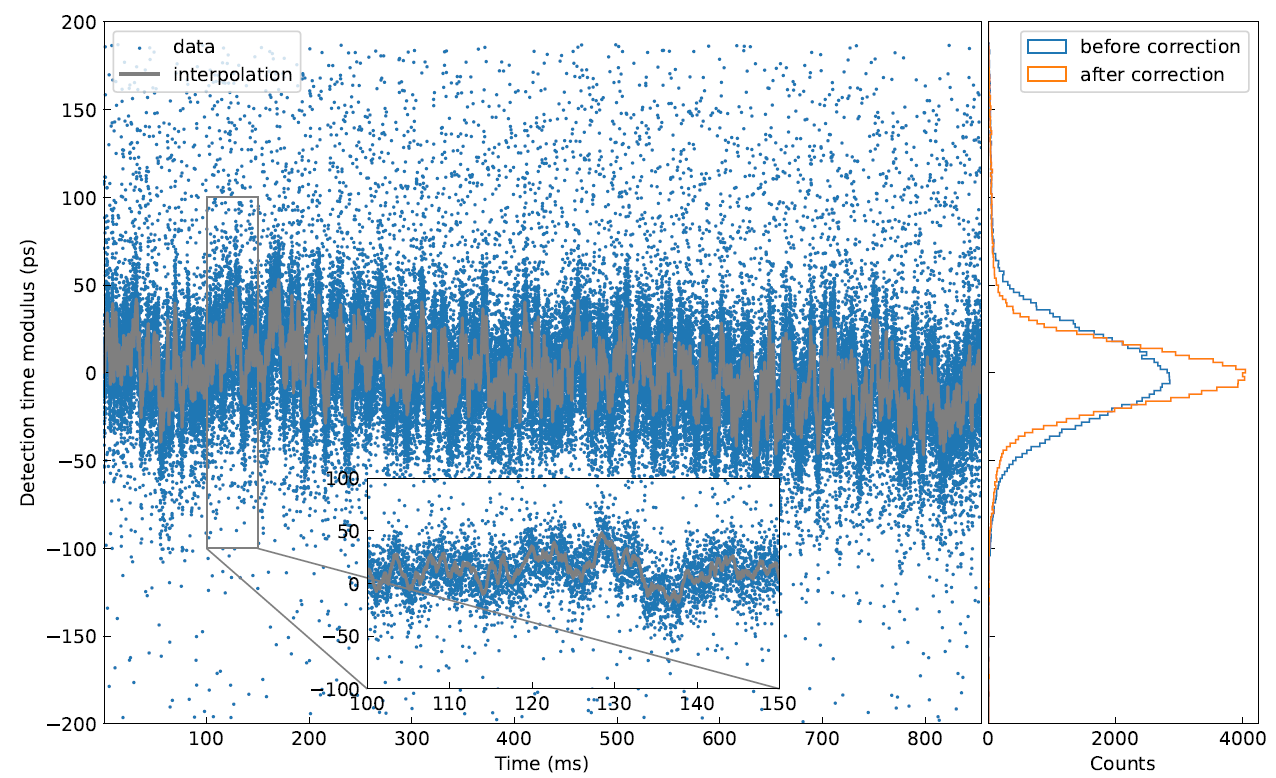}
  \caption{On the left a scatter plot of the detection time moduli of photons belonging to time-bin 0 over time is shown. A fluctuation of the average detection time modulus over time due to imperfect locking between the time-tagger and the pump laser is visible. The grey curve displays a cubic interpolation of the moving average (with a window size equal to 50) of the detection time moduli. On the right, the corresponding histogram produced from the raw time-tags and from the time-tags after correction are shown. Note the reduction of the width of the histogram (the best-fit Gaussian RMS width passes from $24.6$ ps to $16.8$ ps). }
  \label{fig:fluctuations_clock}
\end{figure}

 Our entangled states source was based on the pumping of a SPDC crystal by a mode-locked Ti:Sapphire laser (Coherent MIRA) whose repetition rate is $76$ MHz. We used Single Quantum Eos SNSPD detectors (whose channels RMS nominal jitter ranges from $6.0$ ps to $9.4$ ps) and a Qutools QuTAG time-tagger with jitter upgrade (guaranteeing an RMS jitter lower than $4.5$ ps), but this precision would have been useless without properly locking the time tagger to the fluctuations of the pump laser repetition rate. As the QuTAG can only lock to 10 MHz clock signals, we first electronically derived from the pulses signal at 76 MHz (detected by a fast photo-diode) a 10 MHz signal by exploiting the Clocking Wizard Xilinx® IP core implemented on a Xilinx Zynq®-7000 System-on-Chip. In that way, the histograms displaying the three peaks could be obtained but the three peaks had still consistent overlaps as fitted Gaussians had RMS widths usually ranging between $30$ and $35$ ps (for 1 second of capture). These figures being consistently higher than what was expected, we investigated the evolution of the detection time moduli of photons in a single time-bin over short times scales. We found fluctuations in the average of the modulus of the detection time, as displayed in Fig.~\ref{fig:fluctuations_clock}. These fluctuations can probably be explained by convergence delays or residual errors in the synchronization mechanism between the laser and the time-tagger. The fluctuations could span up to tens of picoseconds over a second and had (detectable) frequency components up to the KHz range. \newline
The detection time moduli $d_i(k)$ of photons generated by pulse $k$ that should belong to time-bin $i \in \left\{1,2,3\right\}$ can thus be modelled as a random process with a fluctuating average $a_i(k)= a_0(k) + i \Delta T$ depending on the accumulated delay between the time-tagger clock and the "laser clock", to which are added several zero-mean random processes: the jitter of the detector $j_{det}$, the jitter of the time-tagger $j_{tag}$, and of finally the intrinsic uncertainty in the detection time of the photons $u_{pho}$ due to their temporal distribution that we will not explicit but is affected by the pump pulse shape (measured through a field autocorrelator to have an intensity profile with best-fit Gaussian RMS width of $4.35\pm 0.12 $ ps), by the random generation positions of the biphoton inside the SPDC crystal (which we expect to introduce a variation in the detection time of up to $3.43$ ps for a KTP crystal of $20$ mm), and by spectral filtering and dispersion effects,
\begin{equation}
    d_i(k) = a_0(k) + i \Delta T + u_{pho} + j_{det} + j_{tag}.
\end{equation}
The fluctuation of the average $a_0(k)$ is the same for all time-bins, so if it were known it could be removed in post-processing. To estimate $a_0(k)$ we designed a simple algorithm that can be run in real-time. We wanted the estimate to be robust against short-term fluctuations in the distribution of detections among the three time-bins, so the algorithm is composed of a first step to guess which events belong to each time-bin and obtain the estimates for the three $a_i(k)$ and a second step in which the three estimates are combined into an estimate of $a_0(k)$.
First, a moving average is computed using the detections in the three time-bins jointly, and it is used to guess with a distance criterion which events belong to the central time-bin and which belong to the lateral ones. Then, for each time-bin we compute again a moving average and interpolate it to all detection times. The differences of the three interpolations of $a_i(k)$ from the expected detection modulus $i\Delta T$ are then combined using a weighted average whose weights are the number of detections used for each interpolation. All tags are then corrected by subtracting this estimate of $a_0(k)$.

%	\newpage
\bibliography{bibliography}

\end{document}